\documentclass[12pt]{article}

\usepackage{graphicx}

\usepackage{epsfig}

\usepackage{bm}
\usepackage{amsmath}

\begin{document}

\title{The nuclear liquid gas phase transition and phase coexistence:A review}


\author{ Philippe CHOMAZ \\
GANIL, BP 5027, 14076 CAEN cedex 5, FRANCE }
 \maketitle


\begin{abstract}
In this talk we will review the different signals of liquid gas phase 
transition in nuclei. From the theoretical side we will first discuss 
the foundations of the concept of equilibrium, phase transition and 
critical behaviors in infinite and finite systems. From the 
experimental point of view we will first recall the evidences for 
some strong modification of the behavior of hot nuclei. Then we will 
review quantitative detailed analysis aiming to evidence phase 
transition, to define its order and phase diagram. Finally, we will 
present a critical discussion of the present status of phase 
transitions in nuclei and we will draw some lines for future 
development of this field.
\end{abstract}

\section{Introduction}

The identification of the various phases of dense matter is one of 
the most important questions of modern nuclear physics. At high 
energy or density one expects to reach a phase in which quarks and 
gluons are deconfined. This plasma of quarks and gluons was the state 
of the matter in the universe shortly after the big-bang prior to its 
condensation in hadrons as schematically shown in the phase diagram 
of figure 1. The amazing progresses in our understanding of this 
phase transition have been extensively discussed during INPC2001 (see 
the present proceeding).

\begin{figure}
 \includegraphics[height=.55\textheight]{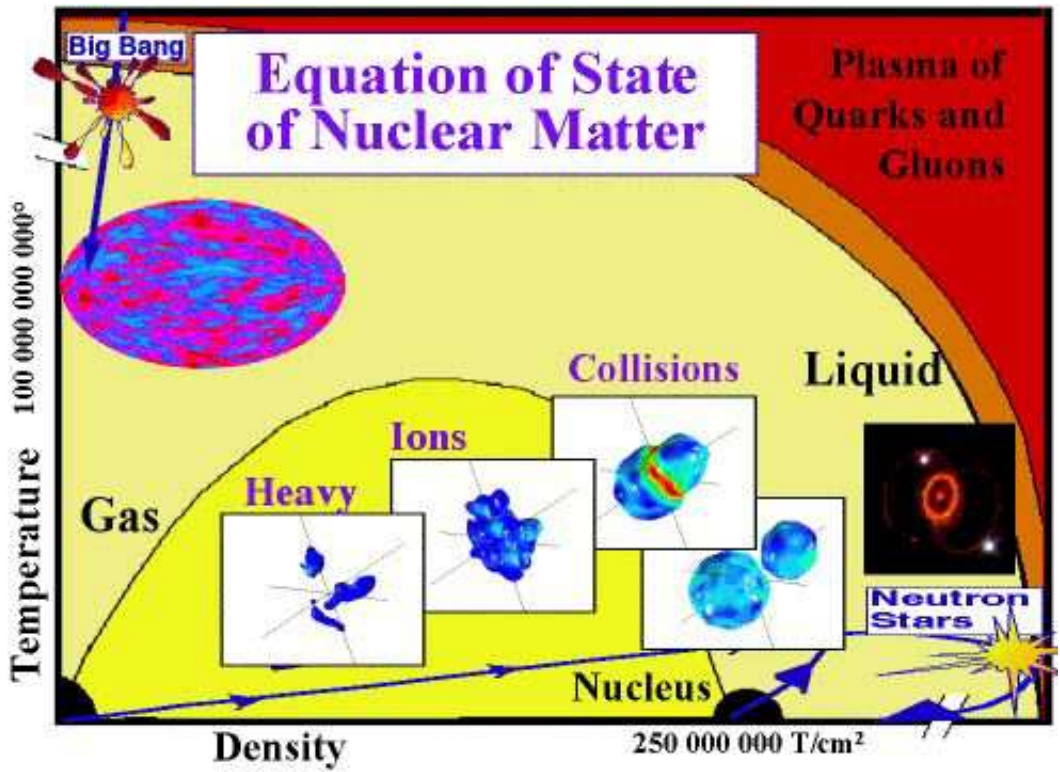}
  \caption{The phase diagram of dense matter. At high energy or density 
  one can see the quark-gluon plasma while around the nuclear matter 
  saturation point the liquid-gas coexistence region is shown. 
  The trajectories followed by different cosmic events are sketched: 
  the big-bang, the collapse of a supernova core in a neutron star 
  and a heavy ion reaction.}
\end{figure}
 
At lower energy one expects a different type of phase transition 
namely the transition between particles and nuclei [1]. This 
condensation is analogous to a liquid-gas phase transition because of 
the resemblance between the nuclear force and a Van der Waals 
interaction with a long range attractive potential and a short range 
repulsive core. This phase transition plays an important role during 
the collapse of supernovae in neutron stars. On earth, physicists 
study it in nuclear reactions such as heavy ion collisions. Fantastic 
progresses have been achieved during a decade and especially in the 
past two years so it is the time to put together all the signals of 
the liquid-gas transition in order to see if they draw a consistent 
picture. As we will see each signal may have its own weak point and 
so is hardly a definitive proof taken individually. However, 
considered as a whole, the ensemble of observations becomes a rather 
strong piece of evidence for a phase transition.

\section{ Something is happening }

As a matter of fact, in the macroscopic world the phase transitions 
manifest themselves by abrupt transformation of the system 
properties. Therefore one may be tempted to first look for rapid 
modifications of physical properties. In the past, many such fast 
transitions have been accumulated [2-6]. Some of them are summarized 
in figure 2. When the excitation energy of the system is increasing 
one observes the disappearance of the heavy residue in the decay 
products of the reaction. For heavy systems, this corresponds to an 
abrupt end of the binary fission. Those channels are in fact replaced 
by an abundant production of fragments, which in turn rapidly 
disappears in favor of the complete vaporization of the composite 
system. Simultaneously, one observes the onset of the radial 
expansion and the associated shortening of the breaking time. All 
these sudden changes in the behavior of the heavy systems recall the 
occurrence of a liquid-gas phase transition. However, they are not 
enough to demonstrate it and a deeper study is called for.  Let us 
first review the theoretical tools needed to understand what is 
happening.

\begin{figure}
 \includegraphics[height=.5\textheight]{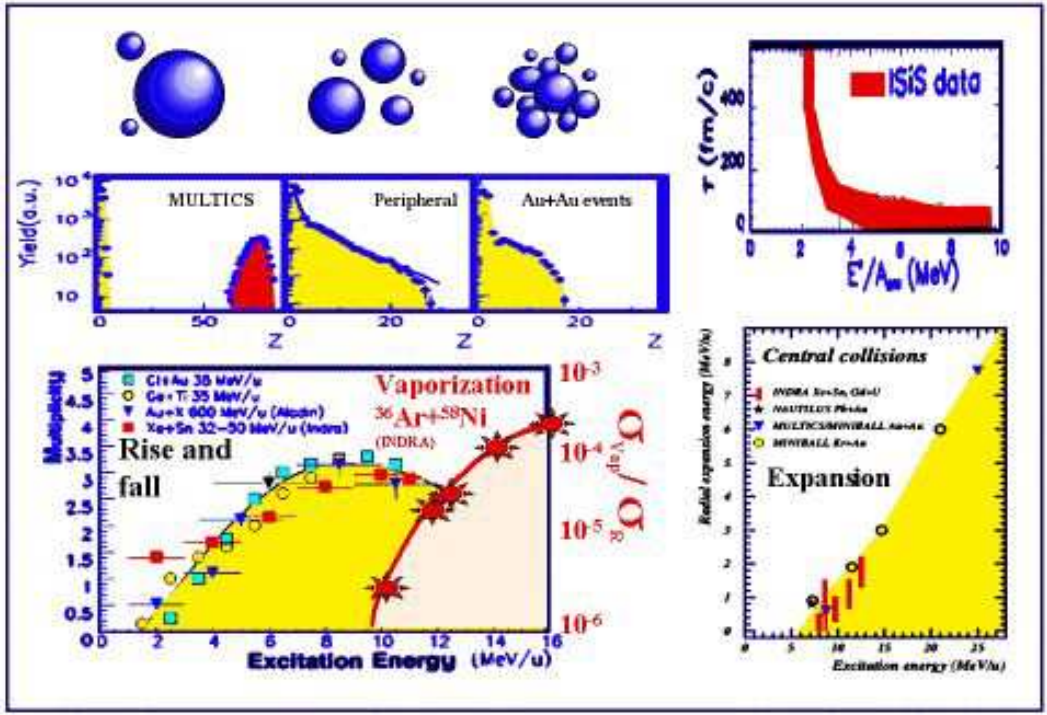}
  \caption{Compilation of many signals of "something is happening"
  when a nuclear system is more and more excited. The left top 
  part is the fragment distribution from ref. [11] showing the 
  disappearance of the heavy fragment. Middle the multiplicity 
  of fragments with Z>2 (IMF) (left scale) demonstrating that 
  the heavy residue is replaced by an abundant production of 
  fragments (the multifragmentation) which are in turn replaced 
  by the vaporization as shown by the cross-section for the 
  total vaporization (right scale). Below, Dalitz plot (the 
  charges of the three largest fragments are  the three 
  distances to the edges) from the Nautilus collaboration 
  showing that the dominant channel for the decay of heavy 
  hot nuclei which is normally the fission (left) is first 
  replaced by the evaporation of IMF (middle) and then by 
  the multifragmentation (right). This is confirmed by the 
  ratio of the ternary and binary fission shown on the right (bottom). 
  Above a compilation of the radial expansion demonstrating the 
  onset of the nuclear explosion around 5-7 MeV excitation energy. 
  Finally, the top right part shows a measure of the fragmentation 
  time, which presents a clear decrease to become almost 
  instantaneous above 3-5 MeV.}
\end{figure}

\section{ Dynamics of the reaction}

\subsection{ Ab initio calculations }

The first idea to control what is happening is to compare experimental 
results to 
dynamical simulation such as transport theory or molecular dynamics 
simulation. The nuclear reaction being very complex the easiest way 
to infer the phase diagram seems to directly study it from the model 
after fitting the various parameters on the multifragmentation data. 
This path towards the nuclear EOS is shown in figure 3. This would be 
the royal way if one would be sure to have the exact description of 
the reaction. However, up to now such a model does not exist and this 
path toward the observation of nuclear phases remains model 
dependent.  

\begin{figure}
 \includegraphics[height=.5\textheight]{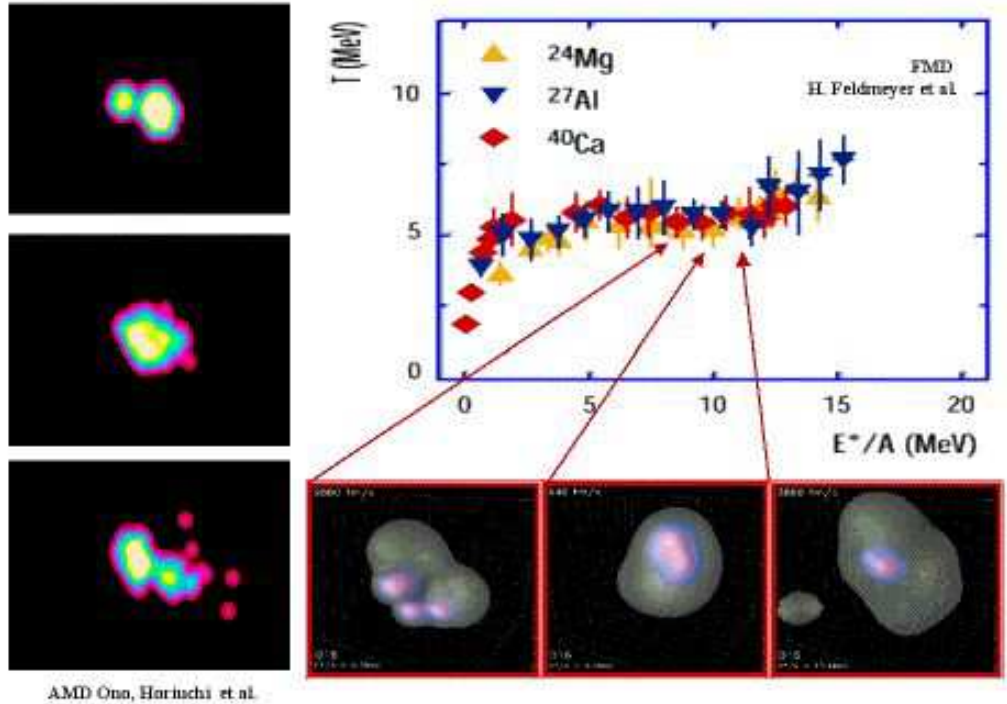}
  \caption{ Molecular dynamics simulations of a collision 
  (left ref. [7]) and of an equilibrium (right [8]). 
  The deduced caloric curve for a confined hot Ca nucleus 
  shows a plateau sign of a 
  liquid-gas phase transition.}
\end{figure}

\subsection{ Dynamics of a phase transition }

Being less confidant in the models, one may try to see direct signals 
of the considered dynamics and, if possible, of the phase transition 
in the data. In particular, many dynamical approaches are predicting 
that during the reaction the system may enter the unstable region of 
the phase diagram and thus can spontaneously undergo a rapid phase 
transition [1,9,10]. This spinodal decomposition is well known in 
many fields of physics. Indeed, deep inside a coexistence region 
uniform systems are generally unstable against fluctuations of the 
associated order parameter. This corresponds to mechanical 
instability for the liquid-gas phase transition or chemical 
instabilities for the mixing of two substances. 

\begin{figure}
 \includegraphics[height=.5\textheight]{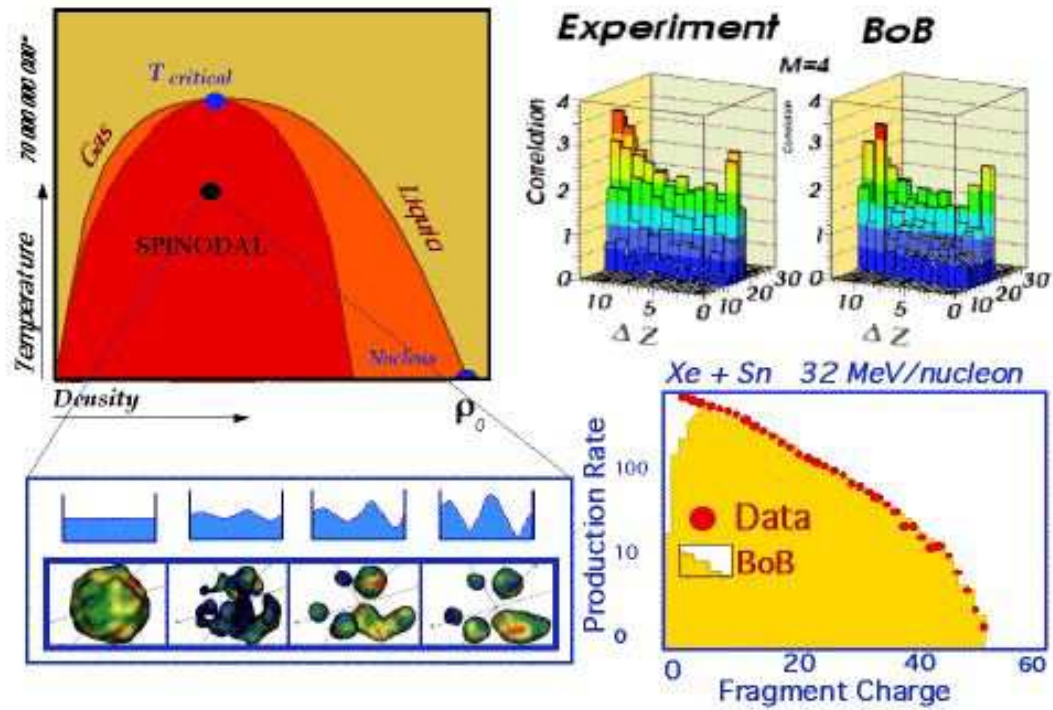}
  \caption{Schematic picture of the phase diagram (left top) 
  showing the spinodal region in the coexistence zone. 
  The small pictures below illustrate the amplification 
  of density fluctuations schematically (above) and (below) 
  as predicted by a stochastic mean-field approximation 
  (Brownian one-Body approach (BoB)). The calculation is 
  compared to the data on the right, for the fragment 
  production measure by the INDRA collaboration for the 
  Xe+Sn reaction at 32 MeV/nucleon (below) and for the 
  associated correlation (above). The correlation corresponds 
  to a sorting of events according to their average and 
  fluctuation of the fragments (Z>2) sizes. The peak at 
  very small fluctuation may be an indication of a favored 
  production of equal size fragments. }
\end{figure}

In particular, it can 
be shown that because of the finiteness of the nuclear forces and of 
the length imposed by the quantum Heisenberg uncertainty principle 
such instabilities presents a favored wave length [10]. This favors 
the breaking of the system in equal size fragments. In the past decade 
a lot of efforts have been devoted to the description of such 
spinodal decomposition in particular using stochastic mean-field 
approaches [10]. It was shown that because of the finiteness of the 
system and of the chaos induced by the non linear regime, which 
follows the early growth of instabilities, such characteristic is 
strongly washed by the end of the dynamics. However, one may try to 
spot some remains of this tendency to split in equal size fragments 
looking at correlations in the fragment distribution. The comparison 
of experimental data with the predictions of stochastic mean-field 
approaches has been recently performed [9]. It was shown that 
fragment distributions, as well kinematical observables and 
correlations are well reproduced by the model. Moreover, a "fossil" 
signal of spinodal decomposition have been reported in the dispersion 
of the fragment sizes [9] (see figure 4).

\subsection{ Dynamics of statistics}

However, it should be noticed that many approaches are able to do an 
almost equally good job as far as the majority of the 
multifragmentation observables are concerned. This is the case of 
many dynamical calculations involving very different approximations 
such as molecular dynamics approaches and stochastic mean-field. Even 
statistical approaches assuming the existence of a freeze-out stage 
in the reaction often fit well the data. This pleads in favor of a 
dominant importance of the phase space irrespectively of the 
considered dynamics. Indeed, the S matrix toward different 
macro-states always contains a micro-state degeneracy factor. Since 
this number varies by huge amount it might well be the dominant factor. 
Moreover, if the dynamics is sufficiently chaotic/mixing a uniform 
population of phase space might be achieved leading to a statistical 
distribution. Therefore one may try to study the reaction using 
statistical mechanics tools.

\begin{figure}
 \includegraphics[height=.3\textheight]{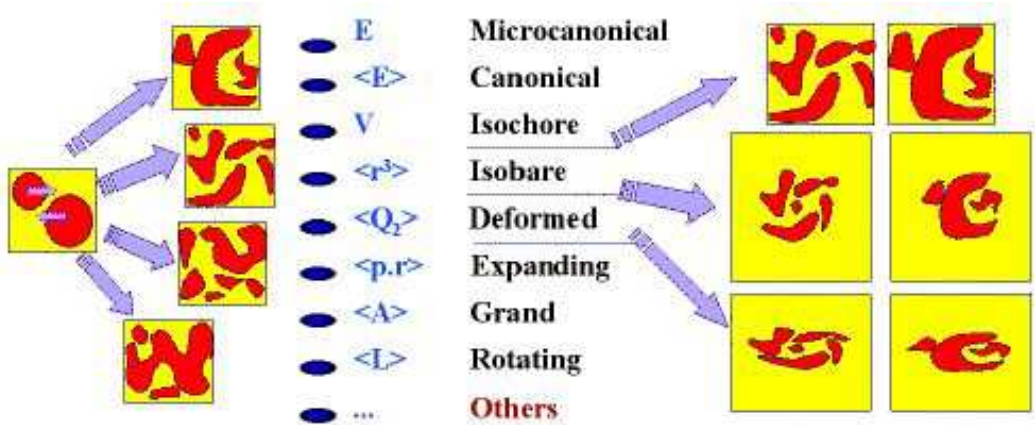}
  \caption{ Schematic illustration of a dynamical evolution of a 
  system into a statistical ensemble when many events are 
  considered and when the dynamics is sufficiently chaotic. 
  Now depending upon the pertinent observable imposed by the 
  dynamics or the event sorting one should describe this set of 
  data using different statistical ensembles some of them being 
  illustrated on the right.     }
\end{figure}

The use of static statistical physics approach to treat a dynamical 
process deserves some additional comments (see figure 5). Indeed, when studying 
nuclear reactions, it is clear that we are facing a dynamical process 
occurring during a finite time. Therefore one should not imagine that 
the statistical physics picture describes an equilibrium in the sense 
of an ergodic evolution of a unique event: the long term behavior (or 
time average) of a long-lived (infinite) system which eventually 
explore the whole phase space. The justification of the use of 
statistical physics to describe transient systems is the fact that an 
ensemble of event, taken at a time, which may fluctuate, from one 
event to another, corresponds to a statistical ensemble. Of course 
the dynamics remains essential in such approach since it determines 
the global variables characterizing the ensemble of events. The 
statistical physics idea is then that these global variables are the 
only important information, the more detailed description being 
governed by randomness. 

It should be stressed that the pertinent information can be the 
result of the dynamics or of the observation process. Indeed, if the 
experimentalists are sorting the events according to a specific 
observable one should take into account this selection of event in 
the theoretical modeling. The application of information theory 
automatically leads to the description of the system as a statistical 
ensemble. Then it should be noticed that many different ensembles can 
be considered depending upon the pertinent information imposed by the 
dynamics and the event sorting. For example, if the dynamics allows 
the energy to fluctuate freely one may try to use the canonical 
ensemble. However, if the energy is conserved or if the events are 
sorted according to their energy one should use a microcanonical 
description. The same discussion can be made for the number of 
particles or the volume. Let us take the example of the volume. If 
the reaction occurs in a fixed volume (i.e. in a box) or if the 
events can be sorted according to their actual volume one should go 
for an isochore ensemble. Conversely if the volume fluctuates from 
event to events in such a way that at most only an average volume can 
be defined one should rather use an isobar picture. Finally one 
should stress that with such a picture of an ensemble of events one 
can also build the statistical description of an evolving system such 
as the rotating or the expanding systems. This only corresponds to 
the introduction of a time odd observable as a constraint in the 
maximization of the entropy.

\section{ Signals of a phase transition}  

In the past years many new analyses have been performed often based 
on novel ideas in order to put in evidence the liquid-gas phase 
transition in nuclei. Let us briefly review the most recent and 
original ones. 

\subsection{ Critical behaviors, the gas side}

Phase transitions are known to be related to critical behaviors and 
to be ruled by universal properties. In particular, at the critical 
point the system presents a fractal structure and so scaling should 
hold. This leads for example to the famous power law shape of the 
fragment size distribution. Moreover, using renormalization group 
argument one can relate the fragment yield and the distance to the 
critical point. Indeed, a yield for a given size at a given distance 
to the critical point is proportional to the yield of a different 
size at a scaled distance. These critical behaviors have been 
identified in many nuclear reactions (see figure 6) [11,12]. The 
inferred critical exponents are in reasonable agreement with those 
expected for the liquid-gas phase transition.  

\begin{figure}
 \includegraphics[height=.5\textheight]{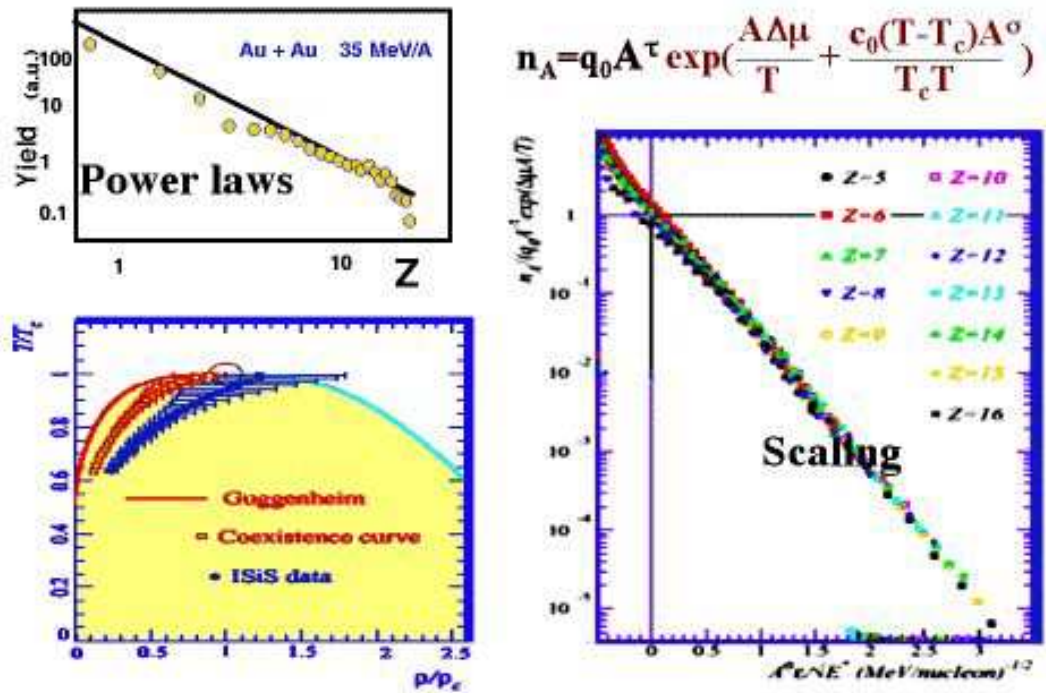}
  \caption{The top-left figure a typical fragment 
  size distribution fitted by a power law indicating 
  the proximity of a critical point. Right the special 
  form of the fisher scaling [14] as tested experimentally 
  in ref. [15]. The resulting scaling function is shown below 
  the equation. The deduced phase diagram is shown on 
  the left (see text).     }
\end{figure}

Recently, it has been proposed [15] to test the specific scaling 
proposed by Fisher [14]. This scaling is based on the simple idea 
that a real gas of interacting particles can be considered as an 
ideal gas of clusters in chemical equilibrium. This can be seen as 
the basis of many multi-fragmentation models except the fact that 
they often take the excluded volume into account. Then the 
population of the various species can be evaluated according to their 
associated free energy. This normally contains a volume and a surface 
term. When the gas is in contact with the liquid the volume term is 
proportional to the difference of chemical potential between the 
liquid and the gas. When the liquid is in equilibrium with the gas 
this term drops off. The surface term is proportional to the surface 
tension which is supposed to decrease and eventually to go to zero as 
we get closer and closer to the critical point. Finally, a 
logarithmic factor is included in the free energy functional in order 
to account for the fractal structure of the fragments at the critical 
point. This topological contribution leads to the famous power law 
distribution of fragment size at the critical point. 

The Fisher scaling gives a specific recipe in order to scale all the 
observed fragment yields on a single curve. This scaling have been 
recently tested with an amazing success on pion-induced fragmentation 
data [15] (see figure 6). Taking the analogy of the perfect gas of 
cluster seriously, the authors of ref. [15] proposed to compute all 
the thermodynamical quantities using the perfect gas equations of 
state. For example, the partial pressure $P_{A}$ induced by the fragments 
A is proportional to $n_{A}T$. The various pre-factors and in particular 
the volume is eliminated by taking the ratio with the thermodynamical 
quantities evaluated at the critical point. This imposes that the 
volume of the considered system is identical to the one of the 
critical point. In such a way, putting the difference of chemical 
potential between liquid and gas to zero one gets the coexistence 
line as shown on figure 6 (see ref [15] for more details). 

The full understanding of this signal of a phase transition needs 
more theoretical and experimental work. First of all, we have 
recently shown that, in very small system, a critical behavior should 
be expected along a line inside the coexistence region [14]. This 
property can also be rather sensitive to the coulomb field. Finally 
the idea that the thermodynamics of a complex real gas can be treated 
as a superposition of clusters ideal gas should be checked (see also 
the chapter about the caloric curve). In this respect the role of the 
total volume of the system and of the volume excluded by each 
fragment should be clarified. However, the observed scaling is 
certainly an interesting signal of a critical behavior.

\subsection{Critical behaviors, the liquid side [16]}

In finite systems not only the small fragment (gas) distribution can 
be studied but also the largest fragment (liquid) can be measured. It 
has been recently discussed that the scaling behavior of this 
distribution may also signal phase transitions [16] of the 
aggregation type. Indeed, within the aggregation scenarios, to which 
belongs the liquid gas phase transition, the order parameter is 
expected to be the size of the largest fragment. Then in ref. [16] it 
is shown that the scaling characteristic of the distribution of order 
parameter should change when passing from one phase to the other. 
More precisely, the order parameter distribution should follow a 
$\Delta$  -scaling with $\Delta$  =1/2 in 
the coexistence (second scaling) 	and $\Delta$  =1 
above the critical point (first scaling). These scaling can be seen 
as very general scaling of the probability distribution of the 
largest fragment size using only two parameters e.g. its maximum (or 
average) and its width. Then, if the maximum is taken as a reference 
and if the standard deviation is used as a unit, the probability, 
correctly normalized to take care of the unit change, might become a 
scale invariant function. For example, with such a procedure all the 
Gaussian distributions collapse onto a single universal curve. Such a 
scaling indicates some relation between the processes associated with 
various distributions, which follow the same scaling. For example, 
all the processes, which fulfill the Laplace central limit 
applicability conditions, pertain to the Gaussian universality class.

\begin{figure}
 \includegraphics[height=.6\textheight]{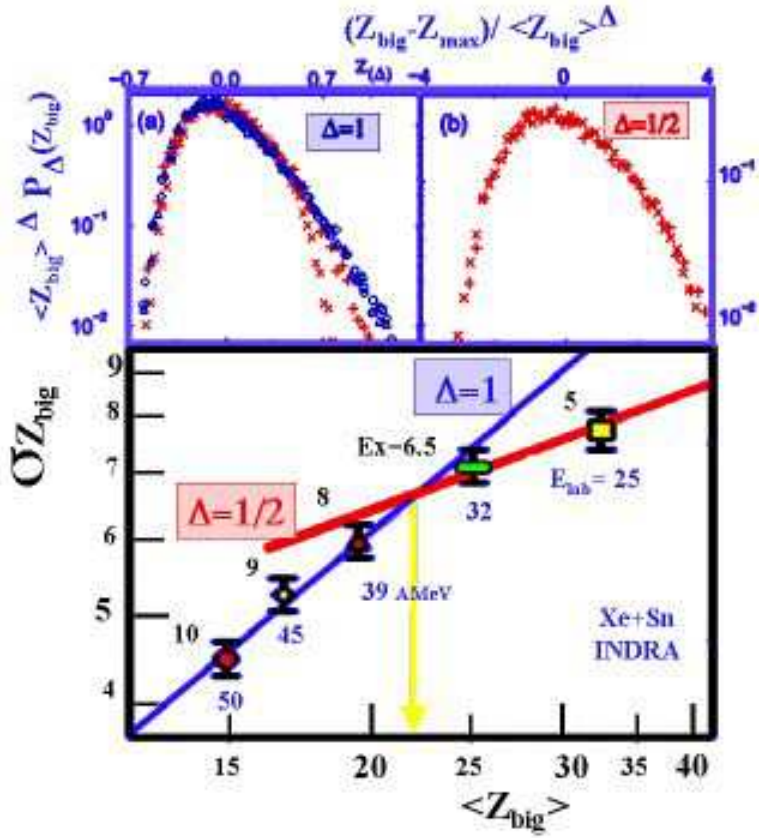}
  \caption{Bottom, the fluctuation of the size of the largest 
  fragment as a function of its average value (both in log scale) 
  as measured by the INDRA collaboration for the central events 
  of the Xe + Sn at 5 incident energies (written bellow the points)
  corresponding to 5 different excitation energies per nucleons of 
  the composite system (written above the points). The two lines 
  show the proposed linear and square root behavior and the arrow 
  shows the assumed point of the phase change. Above are shown 
  two tests of the scaling of the whole distribution using the 
  maximum as the center and the square root of the average as a 
  scale (right) or the average as a scale (left). On the right 
  only the two first energies are shown confirming the second 
  scaling while on the left these energies are shown in a light 
  grey while the three higher energies are in black. Again the 
  black points scale well but not the others demonstrating that 
  only the three highest energies presents a first scaling.  }
\end{figure}

As far as the search for phase transition is concerned, the key point 
is then how the fluctuation changes with the system size or 
equivalently with the maximum or the average of the distribution.  
If the standard deviation scales as a power law of the average then 
the power is called $\Delta$ .  A second scaling ($\Delta$ =1/2) is then the rather 
usual type of fluctuations which goes like in the random walk as the 
square root of the average. The first scaling $\Delta$ =1 corresponds to a 
faster increase of the fluctuations linearly with the mass. The fact 
that it dominates out side the coexistence region can be understood 
since in that case the largest fragment is only the largest one among 
several fragments of comparable small size, i.e. it belongs to the 
gas phase. Therefore, its fluctuations are determined by the fact 
that it should be larger than the second largest fragment. 
Conversely, in the coexistence region the largest fragment is the 
liquid body, which will become infinite at the thermodynamical limit. 
Its fluctuations are ruled by the equilibrium with the gas and so 
look like the one of a random walk process. 

Figure 7 presents the fluctuation of the largest fragment size as a 
function of its averaged value for the central events of the Xe + Sn 
reaction at 5 incident energies [16]. Two behaviors seem to be observed: while 
for the low energy the largest fluctuation seems to behave like the 
root of the average (second scaling), the high energy points rather 
exhibit a linear dependence (first scaling). This observation is 
confirmed by the analysis of scaling of the distribution of sizes 
(figure 7). This would plead in favor of a phase change just above 32 
MeV/nucleon incident energy i.e. about 7MeV excitation energy.  
          
A lot of work is needed before reaching a definite conclusion from 
this signal alone. Indeed, one should study in detail the transition 
region to try to identify how the system goes from one phase to the 
other. Moreover, since the tails of the distribution are important 
for the scaling one should improve the statistics. Also the number of 
energy points should be enlarged as well as other systems analyzed. 
The characterization of "single source" (central) events as a 
function of the bombarding energy should be studied. The role of 
conservation (mass, energy...) should be investigated using for 
example different models. Finally, since this signal is expected for 
many cases from geometrical fragmentation (percolation) to dynamical 
scenarios such as gelation passing by thermodynamical systems such as 
phase transitions (Ising model) one should find other observables in 
order to get a deeper insight into the observed phenomenon.

\subsection{Flattening of the caloric curve} 

In order to get a direct information about the nuclear phase diagram 
and the associated equation of states one should look for direct 
thermodynamical information. In the past year there have been many 
attempts to test if a thermal equilibrium was a reasonable 
approximate description of the fragmenting systems [17-20]. The first 
indication of such an adequacy of a statistical description of a 
freeze-out configuration is given by the amazing success of 
statistical models. However one may look for a direct experimental 
evidence of such an equilibrium at a given stage of the reaction. 

\begin{figure}
 \includegraphics[height=.5\textheight]{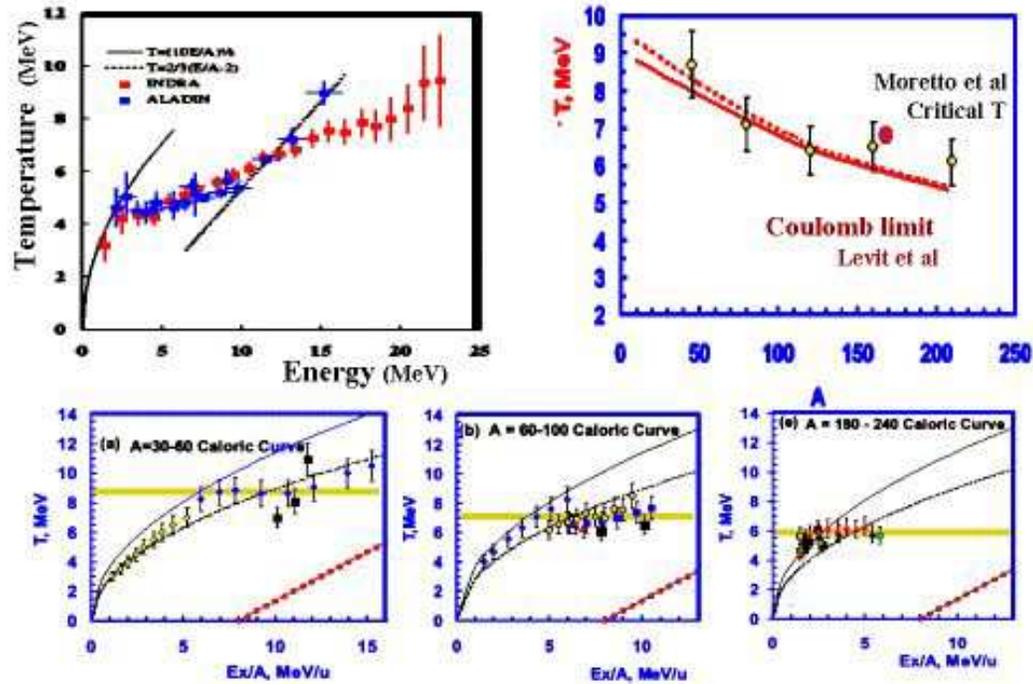}
  \caption{Top-left the caloric curve of the Aladin group [17] 
  compared with the one measured by the indra collaboration [18]. 
  Bottom, caloric curves for different mass of the fragmenting 
  systems coming from many different experiments as analyzed in 
  ref. [20]. The observed mass dependence of the onset of the 
  flattening of the caloric curves is reported in the top-right 
  figure. The point is the critical point discussed in the 
  subsection 1 coming from ref.  [15]. The lines are the limiting 
  temperatures due to the Coulomb instabilities derived in ref. [21].}
\end{figure}

The first result of the Aladin group presenting evidences for a phase 
transition as an almost constant temperature over a broad range of 
energies have triggered a lot of activities (see figure 8). Different 
thermometers were tested, the slope (or the average) of the kinetic 
energies (kinetic temperature), the population of various isotopes 
(chemical temperature) and the ratio between excited states 
population (internal temperature). At the beginning the different 
thermometers and different experiments seemed to not be in agreement.  
However, now, the various observations start to draw a 
coherent picture and the different thermometers agree within the 
experimental error bars as soon we take into account important 
physical effects as:
\begin{itemize}
\item
i) the fact that clusters are not an ideal gas but have a cluster- 
and even state-dependent excluded volume [18];
\item
ii) the fact that the radial expansion affects the kinetic 
temperature both because of the global boost but also because of the 
possible fluctuation of the radial velocity;
\item
iii) the fact that after the freeze-out the fragments should cool 
down leading to a modification of the various population of clusters 
and excited states (side feeding) [19],
\item
iv) the fact that in a hot environment the various states get life 
times which broader their excitation energy affecting their relative 
populations.
\end{itemize}

Recently, it has been shown that the mass of the fragmenting system 
has also an influence on the observed caloric curve [20]. When 
unfolding the effect of the mass the authors of ref. [20] show that 
the caloric curves are less dispersed and present a plateau behavior. 
The temperature of the observed plateau decreases with the mass of 
the fragmenting system. It is interesting to note that the critical 
temperature discussed in the subsection 1 above just lies on the 
curve of the plateau temperature as a function of the mass.
The authors of ref. [20] have compared the observed mass dependence 
of the temperature plateau with the maximum temperature for the 
existence of a charged nucleus in equilibrium with a gas computed in 
ref. [21]. It should be noticed that above this onset of Coulomb 
instabilities only a fragmented system may exist. The amazing 
agreement shown in figure 8 pleads in favor of a relation of the 
observed modification of the trend of the caloric curve with the onset 
of Coulomb instabilities.   

\begin{figure}
 \includegraphics[height=.5\textheight]{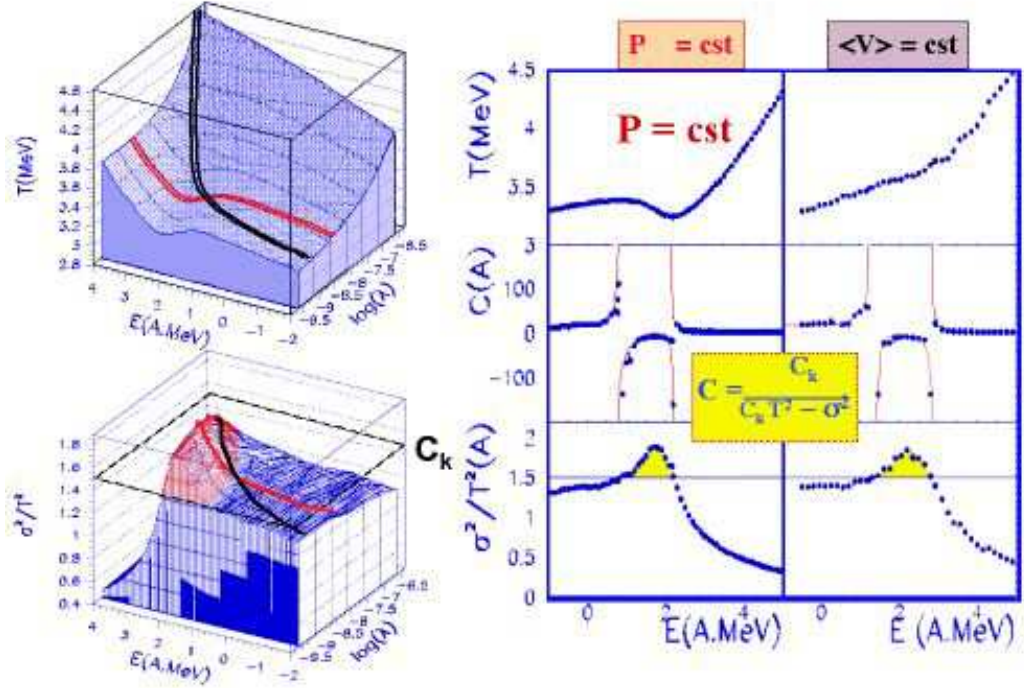}
  \caption{Results from a microcanonical Lattice-gas model with 
  a constraint only on the average volume [22]. Top left the 
  temperature as a function of the energy and the Lagrange 
  multiplier ($\lambda$) which can be seen as a pressure times a 
  temperature. On this curve on the temperature surface are 
  shown two transformations one at constant pressure the other 
  one at constant volume. The corresponding caloric curves are 
  shown on the right. Bottom left the kinetic energy fluctuations 
  for each statistical ensembles characterized by an energy and a 
  Lagrange multiplier. The dotted line marked Ck is the canonical 
  expectation. The curves are the isobar and isochore transformations 
  leading to the two curves shown on the right. Above is the heat 
  capacity of the considered ensemble either computed from the 
  thermodynamics (line) or reconstructed using the kinetic energy 
  fluctuations as expressed in the inserted equation. }
\end{figure}

Of course here too a lot of work remains to be done to strengthen the 
argument. Of particular interest is the behavior of the curve at high 
excitation energy as well as a precise study of the observed 
flattening. It should be stressed that from the theoretical point of 
view in the case of a liquid-gas phase transition one do not expect 
that the phase transition should be marked by a specific behavior of 
the caloric curve. Indeed, the order parameter of the liquid-gas 
phase transition is the density i.e. the volume. Then, the caloric 
curve is not a single curve but a bi-dimensional equation of state 
depending both on energy and volume (or pressure); therefore the 
caloric curve depends upon the actual condition defining the volume 
as illustrated in figure 9 for the lattice-gas model with a volume 
constrained only in average from ref. [22]. However, it should be 
noticed that, while in models the volume can be freely changed, in 
actual nuclear reactions it is determined by the dynamics and cannot 
be controlled but might possibly be measured.

\subsection{Abnormal kinetic energy fluctuations and negative heat 
capacity} 

In ref. [22] it has been proposed to use the kinetic energy 
fluctuation of events sorted in total energy to directly infer 
thermodynamical quantities and more precisely the ensemble heat 
capacity. Indeed, from a classical point of view, for an ensemble at 
constant energy (microcanonical ensemble), the sharing of energy 
between the kinetic and the interaction part should be governed by 
the respective entropies. Then the most probable partition should 
correspond to an equal temperature while the fluctuation should 
depend upon the respective heat capacities. Knowing the kinetic heat 
capacity one can then infer the interaction one and so the total one 
as well as the sum of both (see the equation in figure 9 and ref. 
[22] for more details).

\begin{figure}
 \includegraphics[height=.35\textheight]{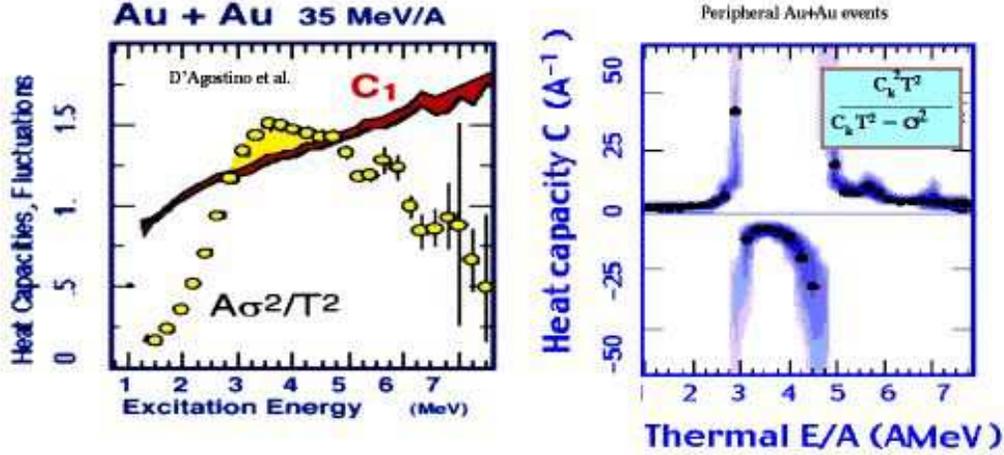}
  \caption{From[27] on the left the kinetic fluctuations 
  divided by the temperature square (dots) compared with 
  the kinetic heat capacity (thick line) on the right the 
  deduced total heat capacity.    }
\end{figure}

\begin{figure}
 \includegraphics[height=.35\textheight]{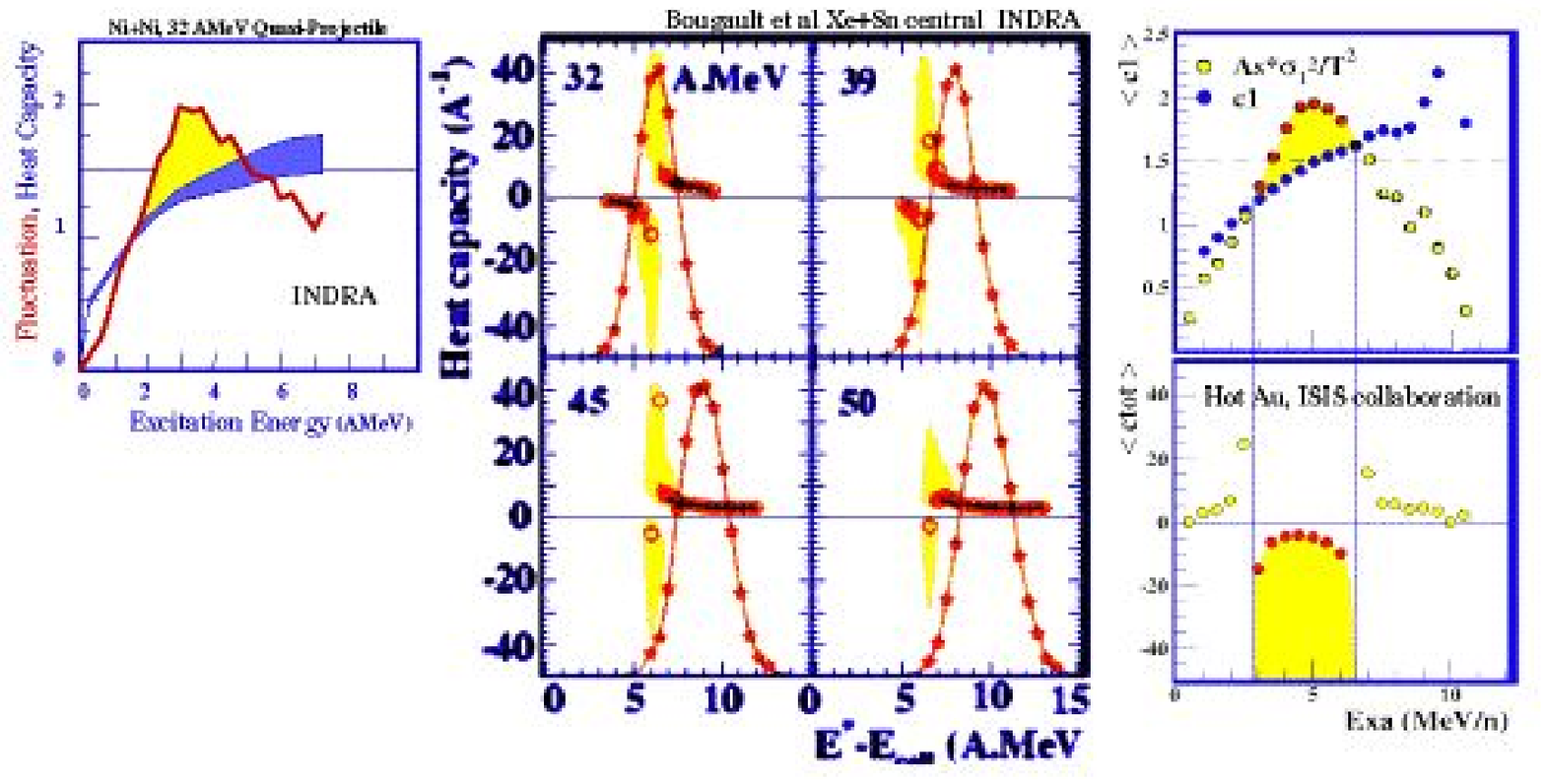}
  \caption{Left the kinetic energy fluctuation divided 
  by the square of the temperature from peripheral Ni+Ni 
  reaction at 32 MeV/A (INDRA collaboration [28]) the thick 
  line is the kinetic energy heat capacity. Middle, reconstructed
  heat capacity from the central events of the Xe+Sn reaction at 
  four different incident energies[28]. Right, top the fluctuation 
  (as in the left curve) and below the deduced heat capacity from 
  the pion reaction on Gold discussed in the subsection 1 [29].     }
\end{figure}

Moreover, it is shown in ref. [22] that in a microcanonical system 
for which the volume is not defined through boundary conditions as it 
is the case for the fragmentation of open systems the phase 
transition is associated with the occurrence of a negative heat 
capacity. 

Negative heat capacities seem impossible from the thermodynamical 
point of view. However, they have been discussed first in the 
astrophysical context [23] of self-gravitating systems. Recently they 
have been pointed out as a possible generic behavior of mesoscopic 
systems undergoing a phase transition, such as in metallic clusters 
[24] and in nuclei [25].  It was recently shown that this concept 
should be extended to inverted curvature of thermodynamical 
potentials as a function of any variables related to the order 
parameter [22]. It should be noticed that the occurrence of negative 
heat capacities has recently be reported for clusters [26]. With the 
nuclear physics results, this is the first experimental evidence for 
such an anomalous behavior.
  
Coming back to the kinetic energy fluctuations as a measure of the 
heat capacity it is shown in ref. [22] that a negative heat capacity 
can be spotted as a microcanonical kinetic energy fluctuation 
becoming larger than the expected canonical limit. Indeed, in the 
equation which relates the heat capacity to the kinetic energy 
fluctuation (see figures 10 and 11) one can see that the denominator is the 
difference between the canonical expectation $C_{k}T^{2}$ and the observed 
fluctuation. When the latter becomes larger than the former the heat 
capacity diverges before becoming negative. 

This signal of a phase transition has been looked for in experiments. 
In such a case an easy splitting of the energy is between the thermal 
excitation and agitation on one side and the partition Q-value plus 
the Coulomb interaction on the other side. The expected canonical 
prediction can be inferred from the relation between the average 
kinetic energy and the temperature since this provides $C_{k}$. Figure 10 
shows the first experimental results of a fluctuation overcoming the 
canonical expectation with the corresponding deduced heat capacity 
for an excited gold nucleus [27]. 

Figure 11 presents an ensemble of results coming from different 
reactions at different energies measured with the Indra [28] and Isis 
[29] apparatuses all presenting abnormally large kinetic energy 
fluctuations and consequently negative heat capacities.

Also for this novel signal of phase transition a lot of work is 
needed both from the experimental and theoretical point of view. 
First many things must be known in order to perform a good total 
energy sorting and to reconstruct the kinetic energy fluctuations at 
freeze out. These reconstructions often need hypotheses such as the 
volume of the freeze-out and the origin of emitted particles. Additional 
measurements to control these hypotheses have to be performed. 
However, kinetic energy fluctuations are a promising way to infer 
thermodynamical properties and to signal phase transitions.       

\subsection{Fractionating distillation of neutron rich matter} 

In ref. [30] it has been discussed that, since the nuclear matter is 
composed of two fluids, neutrons and protons, the liquid-gas phase 
transition should lead to a distillation phenomenon. Indeed, looking 
at the phase diagram as a function of the chemical proportion of 
neutrons and protons y=N/A one can see (figure 12) that, except for 
symmetric nuclear matter, the isospin content of the gas and of the 
liquid is different. The liquid tries to come back to the symmetric 
matter while the gas absorbs the remaining enriched matter. In ref. 
[32], using the lattice-gas model, it was shown that this 
distillation should strongly influence the light fragment production. 
For example, figure 12 shows that for a neutron rich system 
undergoing a phase transition the enrichment in neutron rich isotopes 
(such as the tritium) compared to neutron poor ones (e.g. Helium-3) 
can be much larger than expected on the basis of the N/Z of the 
source.  

\begin{figure}
 \includegraphics[height=.25\textheight]{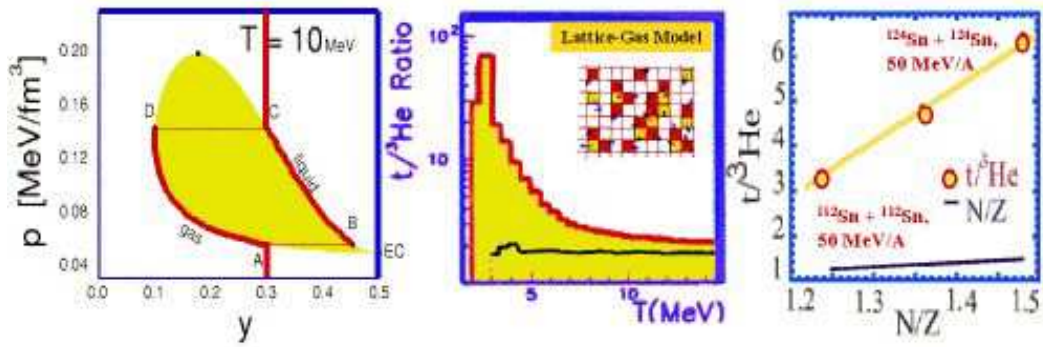}
  \caption{Left the coexistence region at constant temperature 
  in the pressure/neutron-proportion plane  [30]; the thick 
  line is an isothermal transformation. As the pressure reduces 
  the system may reach the coexistence region in C then the liquid 
  goes from C to B, while the gas appears very neutron rich in D 
  and then goes to A. In A the liquid gas transformation is done. 
  Middle, the tritium to helium-3 ratio as a function of the 
  temperature [31]. In the coexistence region below T=6MeV one 
  observes a very strong relative population of neutron rich 
  isotopes. The black line is the result in absence of an isospin 
  dependent interaction term. In such a case the ratio is nothing 
  but the N/Z.  Right, the experimental ratio observed for the 
  central events of different reactions involving two isotopes of
  Sn at 50 MeV/A bombarding energy [32]. The same strong enrichment 
  is observed.    }
\end{figure}

This strong enrichment together with a clear indication of chemical 
equilibrium have been recently reported [33] and interpreted as a 
signature of the fractionation phenomenon. This is a promising avenue 
for the determination of the nuclear equation of state and phase 
diagram. Of particular importance is the isospin dependence of the 
nuclear equation of state, the determination of which can benefit 
from this distillation signal (see also ref. [33] for more details).  

\section{Conclusions and discussion}

\begin{figure}
 \includegraphics[height=.5\textheight]{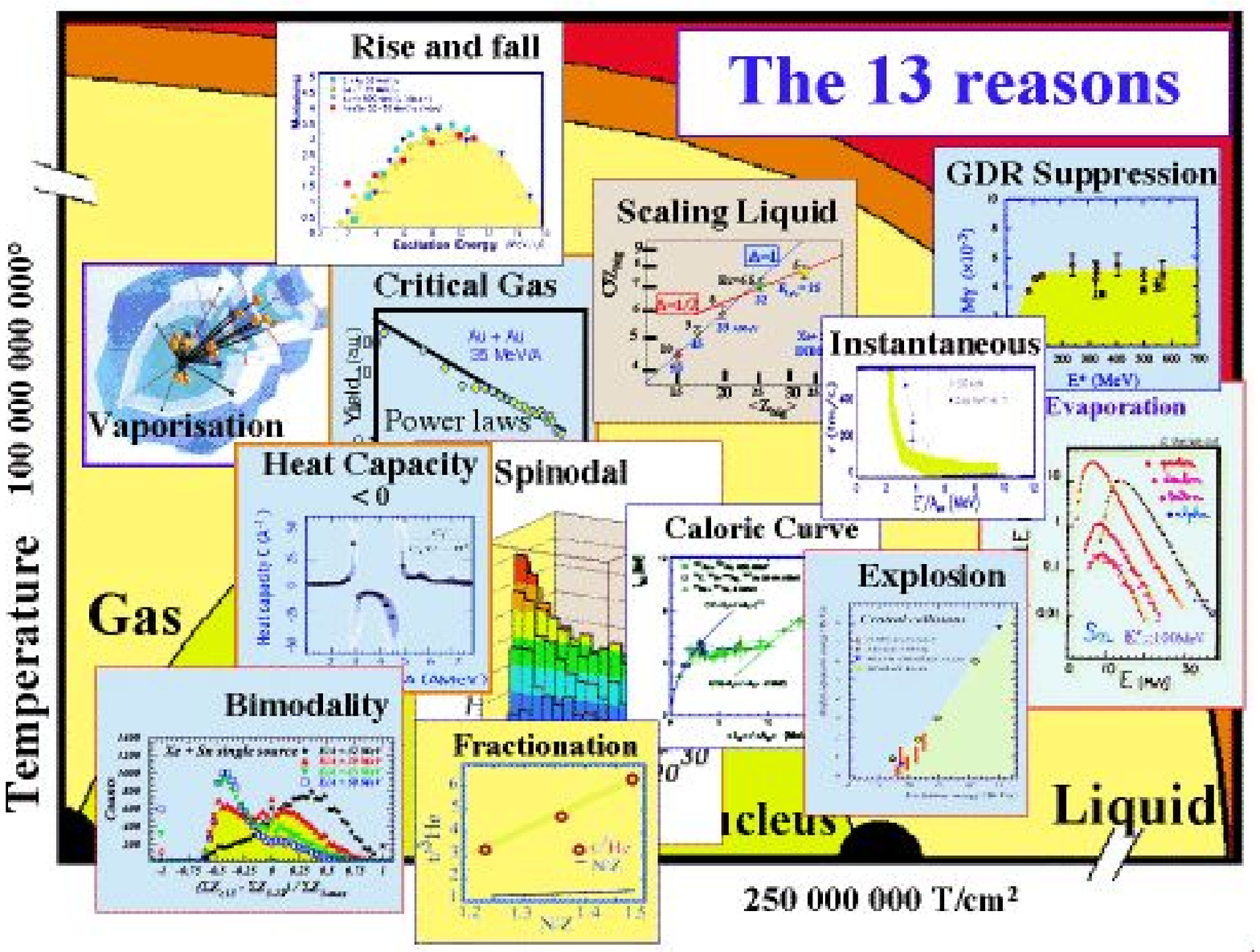}
  \caption{Sketch of the 13 signals of the nuclear liquid 
  gas phase transition (see text).}
\end{figure}
 
\begin{figure}
 \includegraphics[height=.5\textheight]{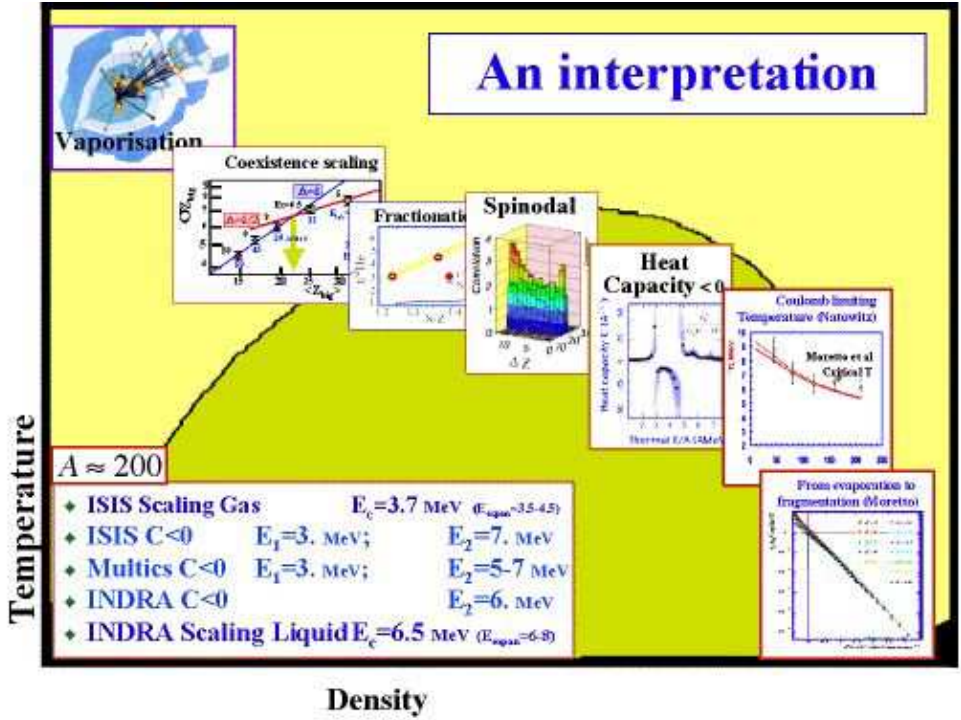}
  \caption{Summary of the recent signals discussed in this 
  review article placed on a possible trajectory of the 
  nuclear systems in the density temperature plane. First 
  at low energy we are in presence of a saturating vapor 
  around the nucleus and one observe the Fisher scaling. 
  Then at a critical temperature the system cannot sustain 
  the Coulomb repulsion and dives into the coexistence zone 
  imposing a mass-dependent bending of the caloric curve. 
  Reaching the spinodal the heat capacity becomes negative
  and some signal of  the spinodal decomposition at the 
  origin of the fragment production may remain. In this 
  region one observe also a fractionation of the neutron 
  rich matter. Then exciting the coexistence it encounters 
  the second divergence of the heat capacity and the end of
  the second scaling of the largest fragment. Then it will 
  gradually go towards a complete vaporization. The table 
  gives, for a mass around 200, the excitation energies at 
  which each signal occurs. One can see that they consistently
  point toward two energies, which might be the entrance and 
  the exit of the coexistence region.    }
\end{figure}

In this review, we have shown that many signals of the liquid-gas 
phase transition have been observed up to now. In figure 13, we have 
collected 13 such evidences:
\begin{itemize}
\item
i) the evaporation of light particles which can be seen as the 
emission of a gas from an isolated piece of matter with no external 
saturating pressure [2-4]
\item
ii) the suppression of the giant dipole vibration marking the end of 
a collective behavior [6] marking the boiling point which for a Sn 
nucleus occurs around 5 MeV temperature (2.5 MeV excitation energy per 
nucleon).
\item
iii) the onset in the same temperature domain of an instantaneous 
fragmentation [2-5]
\item
iv) the bending of the caloric curve again in the same temperature 
domain [17-20]
\item
v) as well as the observation of critical behavior for the gas 
fragments [11-15]
\item
vi) and the onset of the nuclear explosion with a fast radial 
expansion [2-5]
\item
vii) then one observe some fossil signal of a spinodal breaking in 
equal size fragments [9-10]
\item
viii) and at the same time the kinetic energy fluctuation becomes 
abnormal a phenomenon which can be related to the presence of a 
negative heat capacity [22-29]
\item
ix) together with a fractionation which looks like equilibrated [30-33]
\item
x) in this energy domain many bi-modalities of the event distribution 
as a function of various observables are observed as suggested in 
[35] as a signal of phase transition, such as in the difference 
between the mass contained in the fragments and in the light particle 
[34] or in the plane corresponding to the two largest fragment masses 
[2].
\item
xi) then after the rise of the production of large fragments it is 
the time of its fall [2-4]
\item
xii) the scaling of the large fragment mass distribution then passes 
from the second to the first type [16]
\item
xiii) Finnally one observes an equilibrated vaporization [2,3,18]
\end{itemize}

Taken individually each of these 13 signals has its own drawback and 
weakness. However, taken as a whole they start to draw a convicting 
picture of the actual observation of the liquid-gas phase transition 
in nuclei. Not only the reported signals are qualitative they are 
becoming event quantitative (as illustrated on figure 14) allowing to 
think about a real metrology of the nuclear phase diagram. This is a 
good news because a lot remains to be done since not only the 
coexistence zone of the symmetric matter should be measured but also 
the isospin dependence which starts to be experimentally accessible 
thanks to the new radioactive beam factories.



\section{REFERENCES}

\noindent
1.~ 	Siemens P.J., Nature {\bf{305}} 410 (1983),
        Bertsch  G.F. and Siemens P.J., Phys. Lett. B{\bf{126}} 9(1983) 

\noindent
2.~ 	Lopez O., Nucl. Phys. A{\bf 685} 246( 2001)  and refs. there in. 

\noindent
3.~ 	Durand D.,  Tamain B. and Suraud E., "Nuclear Fragmentation" 
        IOP-Publishing 2000 and refs. there in.

\noindent
4.~	 Das Gupta S. et al, nucl-th/0009033 and refs.  there in.

\noindent
5.~ 	Beaulieu L. et al, nucl-ex/0004005, Lefort T. et al, nucl-ex/9910017

\noindent
6.~	Piattelli P. et al, Nucl. Phys. A{\bf 599} 63c(1996) 

\noindent
7.~	Ono A. et al., Phys. Rev. C{\bf 48} 2946(1993) 

\noindent
8.~ 	Schnack J. and Feldmeier H., Phys. Lett. B{\bf 409} 6(1997)
        and Feldmeier H. private comunication

\noindent
9.~ 	Borderie B. et al, nucl-ex/0102015 and 
        Phys. Rev. Lett. {\bf 86} 3252(2001)  

\noindent
10. 	Guarnera A. et al, Phys. Lett. B{\bf 403} 191(1997) , 
        Chomaz Ph.  et al, Phys. Rev. Lett. {\bf 73} 3512(1994) 

\noindent
11. 	D'Agostino M. et al, Nucl. Phys. A{\bf 650} 329(1999); 
        and Elliott J. B. et al, Phys. Rev. Lett. {\bf 85} 1194(2000)  
	
\noindent
12. 	Campi X., Phys. Lett. B{\bf 208} 351(1988) 

\noindent
13. 	 Fisher M.E., Physics {\bf 3} 255(1967) 

\noindent
14. 	Gulminelli F. and Chomaz Ph., Phys. Rev. Lett. {\bf 82} 1402(1999) 

\noindent
15. 	Moretto L.  et al, present proceedings and 
        Elliott J. B. et al, nucl-ex/0104013

\noindent
16. 	Botet R. et al, nucl-ex/0101012 and Phys. Rev. E{\bf 62} 1825(200) 

\noindent
17. 	Pochodzalla J. et al, Phys. Rev. Lett. {\bf 75} 1040(1995) 

\noindent
18. 	Gulminelli F. and Durand D., Nucl. Phys. A{\bf 615} 117(1997)  
        and ref. there in. 

\noindent
19. 	Tsang B. et al, Phys. Rev. C{\bf 53} R1057(1996) 

\noindent
20. 	Natowitz J. et al, nucl-ex/0106016

\noindent
21. 	Bonche P. et al, Nucl. Phys. A{\bf 427} 278(1984)  
        and A{\bf 436} 265(1986) 

\noindent
22. 	Chomaz Ph. and Gulminelli F, Nucl. Phys. A{\bf 647} 153(1999),
        to be published in Phys. Rev. Lett. (2001)
	
\noindent
23. 	 Lynden-Bell M., Physica A{\bf 263} 293(1999); 
         Hauptmannet H. al, Am. J. Phys. {\bf 68} 421(2000)
	 
\noindent
24. 	Labastie P. et al, Phys. Rev. Let.  {\bf 65} 1567(1990)

\noindent
25.     Gross D. H. E., Phys. Rep. {\bf 279} 119(1997) 

\noindent
26. 	Schmidt M. et al, Phys. Rev. Lett. {\bf 86} 1191(2001) 

\noindent
27. 	D'Agostino M. et al,  Phys. Let. B{\bf 473} 219(2000),  
        nucl-ex/0104024 to appear in Nucl. Phys. A (2001)
        and D'Agostino M. private communication

\noindent
28. 	Bougault R. et al, private communication.

\noindent
29. 	Lefort T. et al, private communication. 

\noindent
30. 	Muller H. and Serot B. D., Phys. Rev. C{\bf 52} 2072(1995) 

\noindent
31. 	Chomaz Ph. and Gulminelli F., Phys. Lett. B{\bf 447} 221(1999),
        Pan  J.  and Das Gupta S., Phys. Rev. C{\bf 57} 1839(1998) 

\noindent
32. 	Verde G. et al, Nucl. Phys. A{\bf 681} 299c and 323c(2001), 
        Xu H. S. et al, Phys. Rev. Lett. {\bf 85} 716(2000)  

\noindent
33. 	 Chomaz Ph., Nucl. Phys. A{\bf 681} 199c(2001) 

\noindent
34. 	Borderie B. et al, nucl-ex/0106007

\noindent
35. 	Chomaz Ph.,  Gulminelli F. and Duflot V., Phys. Rev. E (2001) to be published

\end{document}